\documentclass[twocolumn]{aastex62}

\usepackage{graphicx}
\usepackage{array}
\usepackage{amssymb}
\usepackage{float}
\usepackage{color}
\usepackage[utf8]{inputenc} 
\usepackage{amsmath}  
\usepackage{comment}
\usepackage{footnote}
\usepackage{longtable}
\usepackage[caption2]{ccaption}
\usepackage[figuresright]{rotating}
\usepackage{booktabs}
\usepackage[flushleft]{threeparttable}
\usepackage{IEEEtrantools}

\def\rstar {R\textsubscript{$_*$}}
\def\mstar {M\textsubscript{$_*$}}
\def\mp {M\textsubscript{P}}
\def\lbol {L\textsubscript{bol}}
\def\teff {T\textsubscript{eff}}

\newcolumntype{L}{>{\centering\arraybackslash}m{1.5cm}}
\received{XXX}
\revised{XXX}
\accepted{XXX}

\submitjournal{AJ}

\begin{document}

\title{A statistical search for Star-Planet Interaction in the UltraViolet using GALEX}

\correspondingauthor{Gayathri Viswanath}
\email{gayathri.viswanath@astro.su.se}

\author[0000-0003-3250-6236]{Gayathri Viswanath}
\affil{Institutionen f$\ddot{o}$r astronomi, Stockholms universitet, SE-106 91 Stockholm, Sweden}
\affil{Department of Physics \& Electronics, CHRIST (Deemed to be University), Bangalore 560029, India}
\affil{Department of Astronomy and Astrophysics, Tata Institute of Fundamental Research \\
Homi Bhabha Road, Colaba, Mumbai 400005, India}

\author[0000-0002-0554-1151]{Mayank Narang}
\affil{Department of Astronomy and Astrophysics, Tata Institute of Fundamental Research \\
Homi Bhabha Road, Colaba, Mumbai 400005, India}

\author{P Manoj }
\affil{Department of Astronomy and Astrophysics, Tata Institute of Fundamental Research \\
Homi Bhabha Road, Colaba, Mumbai 400005, India}

\author{Blesson Mathew}
\affil{Department of Physics \& Electronics, CHRIST (Deemed to be University), Bangalore 560029, India}

\author{Sreeja S Kartha}
\affil{Department of Physics \& Electronics, CHRIST (Deemed to be University), Bangalore 560029, India}

\begin{abstract}

Most ($\sim$82\%) of the over 4000 confirmed exoplanets known today orbit very close to their host stars, within 0.5 au. Planets at such small orbital distances can result in significant interactions with their host stars, which can induce increased activity levels in them. In this work, we have searched for statistical evidence for Star-Planet Interactions (SPI) in the ultraviolet (UV) using the largest sample of 1355 GALEX detected host stars with confirmed exoplanets and making use of the improved host star parameters from $Gaia$ DR2. From our analysis, we do not find any significant correlation between the UV activity of the host stars and their planetary properties. We further compared the UV properties of planet host stars to that of chromospherically active stars from the RAVE survey. Our results indicate that the enhancement in chromospheric activity of host stars due to star-planet interactions may not be significant enough to reflect in their near and far UV broad band flux.

\end{abstract}


\section{Introduction} 

Exoplanet science has come a long way from the first widely accepted discovery of an exoplanet around a main sequence star in 1995 \citep{MQ1995}. The general architecture and characteristics of planetary systems discovered till date hint at a close relation between planets and their host stars. The mass, radius and orbital distance of planets are a strong function of their host star properties \citep[e.g.,][]{Idalin2005,Cumming2008,Johnson2010,Mordasini2012,Petigura2018,Narang2018,Mulders2018}. Just as the stellar properties influence the properties of their planets, planets can also influence the properties of the host stars. About 82\% of the confirmed exoplanets known today orbit close to their host stars, within 0.5 au. Moreover, about 10\% of the detected planets have mass $>~0.5~\mathrm{M_{J}}$, and orbit at extremely close distances $\le0.1$ au. These are known as `hot Jupiters' and it is postulated that they tend to have stronger magnetic fields than normal Jupiters \citep{Yadav2017,Cauley2019}. At such close distances to their host stars, these massive magnetized planets pave way for interesting interactions with their host stars \citep[e.g.,][]{Cuntz,Cohen2009,Cohen2010,Cohen2011,Lanza2009,Lanza2010,Lanza2011,Pillitteri2010,Pillitteri2014,Saar2004}. 

Star-Planet Interactions (SPI) can enhance and modulate the activity in the upper  atmosphere of the host star \citep{Cuntz}. SPI can be either magnetic or tidal in nature, with the excess stellar activity varying with the planet's orbital period in the former case and with half the orbital period in the latter \citep{Cuntz, Shkolnik2018}. However, confirmations of tidally induced stellar activity enhancements are rare in the literature. The main cause responsible for inducing star-planet interactions seems to be the interactions between the stellar and planetary magnetic fields.

Several attempts have been made in the literature over the years to study SPI \citep[e.g.,][]{Cuntz2002,Shkolnik2003,Shkolnik2005,Shkolnik2008, Kashyap2008, Walker2008,Pagano2009,Lanza2013,Route2019}. \cite{Shkolnik2003} reported the periodic variation of Ca\textsc{ ii} H$~\&$~K lines (considered as indicators of chromospheric activity) in HD 179949, which was in phase with the associated Hot Jupiter's orbital period. \cite{Walker2008} and \cite{Pagano2009} reported the periodic variation in the broad-band optical photometry of $\tau$ Boo and CoRoT-2 respectively. Enhanced stellar activity was reported for HD 17156 in X-rays by \cite{Maggio2015} and for HD 189733 in both X-rays and far-UV by \cite{Pill2011}, which could be associated with a periodic SPI. However, such studies target single sources and require long term observations, often spanning several orbital periods of the planet, which is not always feasible. Moreover, the SPI signatures have been reported to have an intermittent nature due to variations in the stellar magnetic field structure during it's activity cycle \citep{Lanza2008,Lanza2009,Cohen2011}, making the probability of detecting SPI induced activity roughly $\sim75\%$ \citep{Shkolnik2008}. 

In the context of the surge in the discovery of exoplanets recently, a much more convenient and efficient technique to identify SPI signatures could be a statistical approach involving single epoch observations \citep[e.g.,][]{Popp2010,Scharf2010,Popp2011,KB2012,Miller2015,France2018}. Working along these lines, \cite{Kashyap2008} found that main sequence stars with close-in giant planets are on average more X-ray active than those with far-out planets, but these results were later found to not hold for larger samples as demonstrated by \cite{Popp2010} and \cite{Miller2015}, who failed to find any such correlation. \cite{Hartman2010} showed that surface gravity of planets correlated with their host star activity in Ca\textsc{ ii} H$~\&$~K lines for 23 systems with Neptune-mass planets orbiting at $<$~0.1 au.

Theoretically, SPI can be caused by both tidal and magnetic interactions between planets and host stars. However, repeated evidence of SPI signatures have mostly pointed to the origin being magnetic interactions between the active regions of host stars and the magnetosphere of giant planets. Magnetic interactions between the stellar corona and the planetary magnetospheres, which scale as $a^{-2}$, can take place either via magnetic reconnections, propagation of Alfven waves within the stellar wind and generation of electron beams which could potentially strike the base of the stellar corona \citep{Shkolnik2013}. The excess activity caused due to such magnetic SPI tend to vary with the period of the planet's orbit. The extent of the SPI effects depend on the strengths of stellar and planetary magnetic fields as well as the planet's orbital speed relative to the stellar rotation speed \citep{Lanza2012}. Studying these magnetic SPI becomes important mainly because they help detect the planetary magnetic fields, which are difficult to do otherwise \citep{Vidotto2010,Cauley2015,Rogers2017}. Such magnetic SPI mainly manifests itself in the form of enhanced flares in the host stars or hot-spots in the upper stellar atmosphere \citep{RS2000,Lanza2018}. Interactions between the magnetospheres of the star and the planet can cause magnetic reconnections that can produce a beam of charged particles hitting the upper stellar atmosphere and consequently releasing energy in the form of flares. Flares can be emitted in all wavelengths but are particularly prominent in X-rays and UV \citep{Segura2018}. UV radiation due to chromospheric activity is particularly significant in case of low-mass dwarf stars. In addition, the UV continuum flux is predicted to increase in response to increase in the heating rates of the chromosphere and transition regions \citep{Houdebine1996} as well as increase in the stellar rotation rates \citep{Linsky2012}. The FUV region of the stellar spectrum is also rich in spectral lines arising from the chromosphere (e.g., C\textsc{ i}, O\textsc{ i}, C\textsc{ ii}) and transition region (e.g., C\textsc{ iv}, O\textsc{ iv}, O\textsc{ v}; \cite{Linsky2017}) that are in the GALEX pass band. Considering the possibility of detecting enhancements in stellar UV activity using GALEX, \cite{Shkolnik2013} investigated the SPI effects on the UV activity of about 272 FGK host stars detected by GALEX by searching for statistical correlations with planetary properties \mp, $a$ and \mp/$a$. While they did not find any correlation for Near Ultra Violet (NUV) activity, they report  tentative evidence of SPI in the Far Ultra Violet (FUV) activity, with a correlation at a level of 1.8$\sigma$ for the radial velocity detected planets and at a level of 2.3$\sigma$ for the sample containing both the radial velocity and transit detected planets. However, since 2013, the number of confirmed exoplanet detections have increased multifold and these results need to be carefully revisited using the much larger sample of exoplanets.

In our work, we use the final data release of GALEX \citep{Morrissey2007}, GR6/GR7, to obtain the UV flux of host stars along with their $Gaia$ DR2 stellar properties to search for statistical correlations between the UV activity of the largest sample of host stars with confirmed planets and their respective planetary properties. In Sect. 2, we describe the sample used for this work. In Sect. 3, we explain the results from this study. A discussion about the obtained results is presented in Sect. 4 and the major results from this study is summarized in Sect. 5.

\section{Sample and Data Analysis}

The sample used for this study is obtained from the Confirmed Planets Table at the NASA Exoplanet Archive (NEA), consisting of 3885 planets and 2900 host stars (as on January 17, 2019). We cross matched this sample of host stars with $Gaia$ DR2 using their coordinates in 2015.5 epoch via the Mikulski Archive for Space Telescopes (MAST) to retrieve accurate measurements of stellar parameters along with their uncertainties. We used an initial search radius of $10\arcsec$ but the search radius was increased to $30\arcsec$ for cases when a match was not found within $10\arcsec$. Besides the 1431 sources which returned a single {\it Gaia} DR2 match, we found multiple {\it Gaia} DR2 detections for 1451 sources, from which the source of interest was identified using their $||G-V||$ magnitude as well as their distance from the queried coordinate. This method is outlined in detail in Narang et al. (2020, in prep). Thus $Gaia$ DR2 data were obtained for 2882 host stars, among which 2829 had parallax values available. The distance of these sources were obtained from \cite{BJ2018}. 

To find the host stars detected in UV, these 2829 host stars were then cross matched with the final release of GALEX data, GALEX GR6/GR7. We used a cross match radius of 6\arcsec~and accessed the data using the GALEX Merged Catalog (MCAT) (via GalexView). To maintain homogeneity in the sample, we only chose sources detected in the GALEX All-Sky Imaging  Surveys (AIS). For 1206 host stars, the cross match returned a single GALEX match. For the 149 sources for which GALEX returned multiple matches within 6\arcsec, the true match was identified using the criteria from \cite{Bianchi2017} according to which the match with the longest exposure time, or in case of equal exposure times, the one with the shortest separation from the field centre was selected. Through this process we identified true GALEX matches to 149 sources from among 463 multiple GALEX detections. Thus, we found GALEX GR6/GR7 data for 1355 stars. 

Of these 1355 stars, 1328 were detected in the NUV band and 302 were detected in the FUV band. We further filtered out sources which raised an extraction or artifact\footnote{http://galex.stsci.edu/GR6/?page=ddfaq\#6} flag of $>$ 0 \citep{Bianchi2017, Shkolnik2013}, to avoid any contribution to the flux due to possible window/detector/dichroic reflection, other UV sources in the vicinity of the source or ghost images. Following this, we obtained a sample of 593 NUV detected host stars with 761 planets and 264 FUV detected host stars with 335 planets. The FUV sample of host stars have no source brighter than magnitude 15, but in the NUV sample $\sim$25\% are brighter than magnitude 15, with $\sim$10\% brighter than 13.8. These sources would face the issue of saturation, ie., their flux is underestimated by GALEX \citep{Morrissey2007}. However, at 13.8 magnitude the flux only changes by 10\%, so this reduction is not significant for our sample as it only affects a small fraction of the sample by a small amount. Further, there is a large scatter in the calibration data used by \cite{Morrissey2007}, indicating that it is difficult to accurately correct for this flux saturation. Hence, we will use the GALEX fluxes for these sources in our work, assuming that the effect of saturation in our sample is not significant.

We obtained the planetary properties of our host stars from NEA. For those planets with data missing from the `Confirmed Planets Table' at NEA, data was taken from the Composite Planet Data. We only retained those sources in our sample that have the observed stellar (\teff, \lbol, \rstar, parallax, \mstar) and planetary parameters (orbital period, \mp) with $\geq3\sigma$ confidence. This gave us a sample of 213 stars detected in NUV with 255 planets and 153 stars detected in FUV with 200 planets. The stellar surface gravity values, log~\textsl{g}, were available for 212 stars having NUV data and 153 stars having FUV data. Using the log~\textsl{g} values, we further filtered out the evolved stars from our sample based on the criterion from \cite{ciardi2011}. This recognizes an evolved star as the one in the surface gravity range:

\begin{equation}
\mathrm{log~\textsl{g}} < \left\{ \,
\begin{IEEEeqnarraybox}[][c]{l?s}
\IEEEstrut
3.5 & if $\mathrm{T_{eff} (K)} \geq 6000$ \\
4.0 & if $\mathrm{T_{eff}(K)} \leq 4250$ \\
5.2 - (2.8 \times 10^{-4}\,\mathrm{T_{eff}}) & if $4250<  \mathrm{T_{eff}(K)}< 6000$.
\IEEEstrut
\end{IEEEeqnarraybox}
\right.
\end{equation}

Thus, the final sample for our analysis consists of 178 main sequence dwarfs with 215 planets detected in NUV (see Table 1), and 123 dwarfs with 166 planets detected in FUV (see Table 2).

\begin{longtable*}[c]{|l|L|L|L|L|L|L|L|L|}
\caption{Parameters for the 215 planets around 178 main sequence stars detected in GALEX NUV band. The entire table is available in the electronic form.}
\\
\hline
No. & \textbf{Host Star} & $\mathbf{T_{eff}\,(K)}$ & $\mathbf{M_*\,(M_{\odot})}$ & $\mathbf{log\,\frac{L_{NUV}}{L_{bol}}}$ & \textbf{Planet} & $\mathbf{M_P\,(M_J)}$ & \textbf{Orbital distance (au)} & \textbf{Discovery Method} \\ [1ex]\\ \hline
\endfirsthead
\multicolumn{9}{c}%
{{\bfseries Table \thetable\ continued from previous page}} \\
\endhead
1 & KELT-12 & 5994.6 $\pm$ 139.28 & 1.59 $\pm$ 0.08 & -1.992 $\pm$ 0.0002 & KELT-12 b & 0.95 $\pm$ 0.14 & 0.067 $\pm$ 0.0011 & Transit \\ \hline
2 & WASP-136 & 6400.4 $\pm$ 232.17 & 1.41 $\pm$ 0.07 & -1.521 $\pm$ 0.0006 & WASP-136 b & 1.51 $\pm$ 0.08 & 0.066 $\pm$ 0.0011 & Transit \\ \hline
3 & WASP-159 & 5909.0 $\pm$ 76.12 & 1.41 $\pm$ 0.12 & -2.052 $\pm$ 0.0003 & WASP-159 b & 0.55 $\pm$ 0.08 & 0.054 $\pm$ 0.0015 & Transit \\ \hline
4 & HATS-26 & 5923.2 $\pm$ 185.88 & 1.3 $\pm$ 0.085 & -1.535 $\pm$ 0.0011 & HATS-26 b & 0.65 $\pm$ 0.076 & 0.047 $\pm$ 0.001 & Transit \\ \hline
5 & Kepler-5 & 5718.3 $\pm$ 354.39 & 1.37 $\pm$ 0.05 & -2.019 $\pm$ 0.0005 & Kepler-5 b & 2.11 $\pm$ 0.076 & 0.05 $\pm$ 0.0006 & Transit \\ \hline
6 & HAT-P-65 & 5648.9 $\pm$ 223.97 & 1.21 $\pm$ 0.05 & -2.376 $\pm$ 0.0002 & HAT-P-65 b & 0.53 $\pm$ 0.083 & 0.039 $\pm$ 0.0005 & Transit \\ \hline
7 & HAT-P-66 & 6440.0 $\pm$ 811.3 & 1.25 $\pm$ 0.08 & -1.489 $\pm$ 0.0012 & HAT-P-66 b & 0.78 $\pm$ 0.057 & 0.044 $\pm$ 0.0009 & Transit \\ \hline
8 & WASP-63 & 5475.3 $\pm$ 228.64 & 1.28 $\pm$ 0.42 & -2.547 $\pm$ 0.0 & WASP-63 b & 0.37 $\pm$ 0.09 & 0.057 $\pm$ 0.0062 & Transit \\ \hline
9 & HD 154857 & 5582.5 $\pm$ 76.95 & 1.96 $\pm$ 0.12 & -2.463 $\pm$ 0.0004 & HD 154857 c & 2.58 $\pm$ 0.16 & 5.558 $\pm$ 0.1598 & Radial Velocity \\ \hline
10 & HD 154857 & 5582.5 $\pm$ 76.95 & 1.96 $\pm$ 0.12 & -2.463 $\pm$ 0.0004 & HD 154857 b & 2.45 $\pm$ 0.11 & 1.342 $\pm$ 0.0274 & Radial Velocity \\ \hline
\end{longtable*}

\begin{longtable*}[c]{|l|L|L|L|L|L|L|L|L|}
\caption{Parameters for the 166 planets around 123 main sequence stars detected in GALEX FUV band. The entire table is available in the electronic form.}
\\
\hline
No. & \textbf{Host Star} & $\mathbf{T_{eff}\,(K)}$ & $\mathbf{M_*\,(M_{\odot})}$ & $\mathbf{log\,\frac{L_{FUV}}{L_{bol}}}$ & \textbf{Planet} & $\mathbf{M_P\,(M_J)}$ & \textbf{Orbital distance (au)} & \textbf{Discovery Method} \\ [1ex]\\ \hline
\endfirsthead
\multicolumn{9}{c}%
{{\bfseries Table \thetable\ continued from previous page}} \\
\endhead
1 & HD 106270 & 5562.0 $\pm$ 57.0 & 1.33 $\pm$ 0.05 & -5.177 $\pm$ 0.0 & HD 106270 b & 10.13 $\pm$ 0.27 & 3.268 $\pm$ 0.0449 & Radial Velocity \\ \hline
2 & HD 88133 & 5413.7 $\pm$ 96.38 & 1.26 $\pm$ 0.25 & -5.57 $\pm$ 0.0 & HD 88133 b & 0.28 $\pm$ 0.006 & 0.048 $\pm$ 0.0032 & Radial Velocity \\ \hline
3 & KELT-12 & 5994.6 $\pm$ 139.28 & 1.59 $\pm$ 0.08 & -4.435 $\pm$ 0.0 & KELT-12 b & 0.95 $\pm$ 0.14 & 0.067 $\pm$ 0.0011 & Transit \\ \hline
4 & WASP-136 & 6400.4 $\pm$ 232.17 & 1.41 $\pm$ 0.07 & -3.785 $\pm$ 0.0 & WASP-136 b & 1.51 $\pm$ 0.08 & 0.066 $\pm$ 0.0011 & Transit \\ \hline
5 & XO-3 & 6885.3 $\pm$ 233.68 & 0.58 $\pm$ 0.14 & -3.149 $\pm$ 0.0 & XO-3 b & 7.29 $\pm$ 1.19 & 0.035 $\pm$ 0.0028 & Transit \\ \hline
6 & 70 Vir & 6245.0 $\pm$ 603.75 & 1.14 $\pm$ 0.08 & -4.673 $\pm$ 0.0 & 70 Vir b & 7.42 $\pm$ 0.057 & 0.486 $\pm$ 0.0114 & Radial Velocity \\ \hline
7 & HD 10697 & 5654.0 $\pm$ 45.76 & 1.11 $\pm$ 0.02 & -2.972 $\pm$ 0.0006 & HD 10697 b & 6.38 $\pm$ 0.078 & 2.116 $\pm$ 0.0127 & Radial Velocity \\ \hline
8 & HD 163607 & 5524.2 $\pm$ 55.31 & 1.12 $\pm$ 0.16 & -5.308 $\pm$ 0.0 & HD 163607 c & 2.2 $\pm$ 0.037 & 2.373 $\pm$ 0.113 & Radial Velocity \\ \hline
9 & HD 163607 & 5524.2 $\pm$ 55.31 & 1.12 $\pm$ 0.16 & -5.308 $\pm$ 0.0 & HD 163607 b & 0.78 $\pm$ 0.01 & 0.361 $\pm$ 0.0172 & Radial Velocity \\ \hline
10 & HD 222155 & 5720.0 $\pm$ 44.25 & 1.21 $\pm$ 0.1 & -5.017 $\pm$ 0.0 & HD 222155 b & 2.12 $\pm$ 0.5 & 5.221 $\pm$ 0.4579 & Radial Velocity \\ \hline
\end{longtable*}

\section{Results}

For strong signatures of SPI, the planet needs to be massive and close to the host star such that the magnetic fields of the planet and the star can interact. In addition to the planet's orbital distance $a$, the ratio of the planet mass to its semi-major axis, \mp/$a$, has also been used in the literature as a planetary parameter whose variation with stellar activity could be used to search for SPI \citep[e.g.,][]{Popp2010, Shkolnik2013, France2018}.Thus, we analyzed the behaviour of the latest GALEX NUV and FUV luminosities of the largest sample of host stars currently available against their planetary properties $a$ and \mp/$a$ to look for signatures of possible Star-Planet Interaction statistically.

\subsection{Comparing the distributions of planet hosting and non-planet hosting stars}

In order to compare the activity of planet hosting stars in our sample to that of non-planet hosting stars, we retrieved $Gaia$ DR2 data for a sample of 61438 main sequence stars within 100 pc and having \teff~ and \lbol~ values listed in $Gaia$ DR2 with a 3$\sigma$ or higher significance. Using the median \lbol~for each spectral type from \cite{PM2012}, we selected a thin strip of 11166 stars along the main sequence line in the HR diagram. Using their J2000 coordinates, we cross matched these main sequence stars with GALEX GR6/GR7 with a search radius of 6\arcsec. This gave us a sample of 5477 $Gaia$ DR2 listed main sequence stars within 100 pc which were detected in GALEX AIS survey. Out of these, all 5477 had NUV flux and 1672 had FUV flux listed in GALEX. After removing sources with bad flags (artifact and extraction flag $>$0) and those with flux measured below 5$\sigma$, we arrived at a sample of 2202 stars with NUV flux and 790 stars with FUV flux values. To make sure that none of these sources are stars that host any known planets, we then removed all those sources from this sample that came within 6\arcsec~of our main sequence planet hosting stars sample. Thus, we obtained a control sample of 2197 NUV and 783 FUV $Gaia$ DR2 main sequence stars within 100 pc that are not known to host planets. 
\begin{figure}[ht!]
\centering
\includegraphics[width=1.15\linewidth]{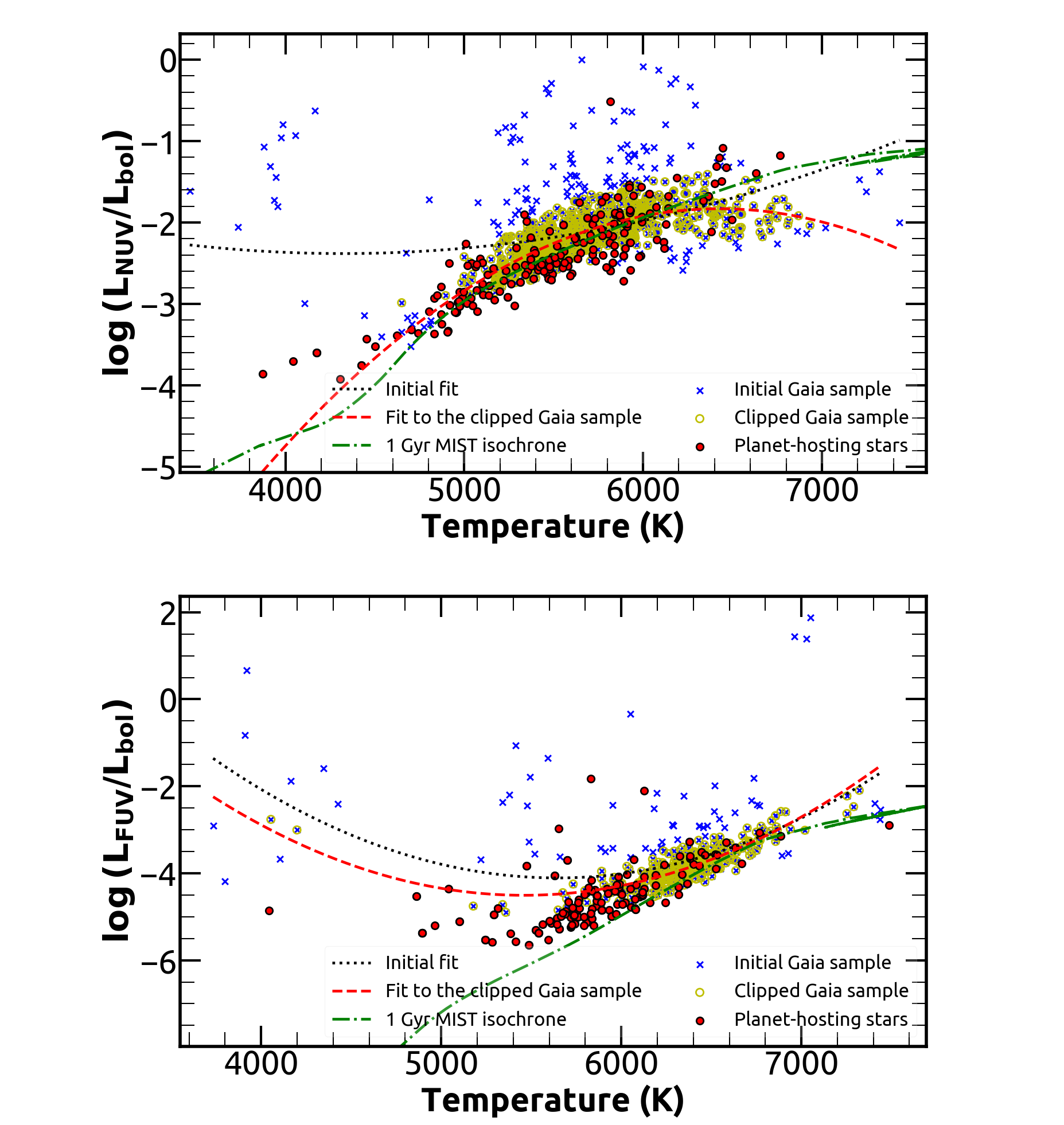}
\caption{The sample of planet-hosting stars detected in NUV (top panel) and FUV (bottom panel) with {\it Gaia} DR2 properties. The red solid circles represent the host stars. The blue crosses represent the {\it Gaia} main sequence sample within 100 pc and the yellow circles represent the 3 sigma clipped subsample of the {\it Gaia} sample. The black and red dashed lines represent the fit to the original and the 3 sigma clipped sample of {\it Gaia}-detected non-planet hosting dwarfs within 100 pc respectively. The green dashed line represents the 1 Gyr MIST isochrone for dwarfs.} \label{Figure1}
\end{figure}
Figure \ref{Figure1} shows the final sample of planet hosting stars detected in NUV and FUV with {\it Gaia} DR2 properties, along with the non-planet hosting stars sample. The median uncertainty in the \teff~ is 105 (60, 158) K for the NUV sample and 63 (45, 102) K for the FUV sample respectively, where the values in the parenthesis represent the lower and the upper quartiles. The median error in the figure for $\mathrm{L_{NUV}/L_{bol}}$ is 3.5 $\times 10^{-4}$ (1.7 $\times 10^{-4}$, $10^{-3}$) and for $\mathrm{L_{FUV}/L_{bol}}$ is 4.2 $\times 10^{-6}$ (2.2 $\times 10^{-6}$, 1.4 $\times 10^{-5}$). Figure \ref{Figure1} also shows the behaviour of the fractional UV luminosity of a typical 1 Gyr old main sequence star with temperature as predicted by the MIST model, along with a polynomial fit to the observed distribution of $Gaia$ main sequence non-planet hosting sample (both explained in detail in section 3.2). The divergence of the MIST model from the observed values at lower \teff~ as seen in the figure is expected since MIST models only account for the photospheric emission and not the emission from chromosphere and transition region of the star. But at lower temperatures, photospheric emission contributes very little to the NUV and FUV flux from the star, which causes MIST model to make bad predictions in the lower \teff~ regime. However, from  Figure \ref{Figure1} one can see that there is no clear distinction between the distributions of planet hosting and non-planet hosting stars.

Although we have removed all known planet hosting stars from our control sample, stars with yet undetected planets could still contaminate the sample. However, from Figure \ref{Figure1}, we see that the predominant spectral type of the control sample is F-G-K type stars, for which the occurrence rate of hot Jupiters which are responsible for causing SPI-induced activity enhancement is as low as 0.7 hot Jupiters per 100 stars (This value has been obtained based on the treatment in \cite{Narang2018} and is similar to what is presented in \cite{Hsu19}). This is not high enough to alter the total distribution of our control sample and hence our general conclusion will remain unaffected even if such a contamination is present.

\begin{figure*}[ht!]
\centering
\includegraphics[width=1.0\linewidth]{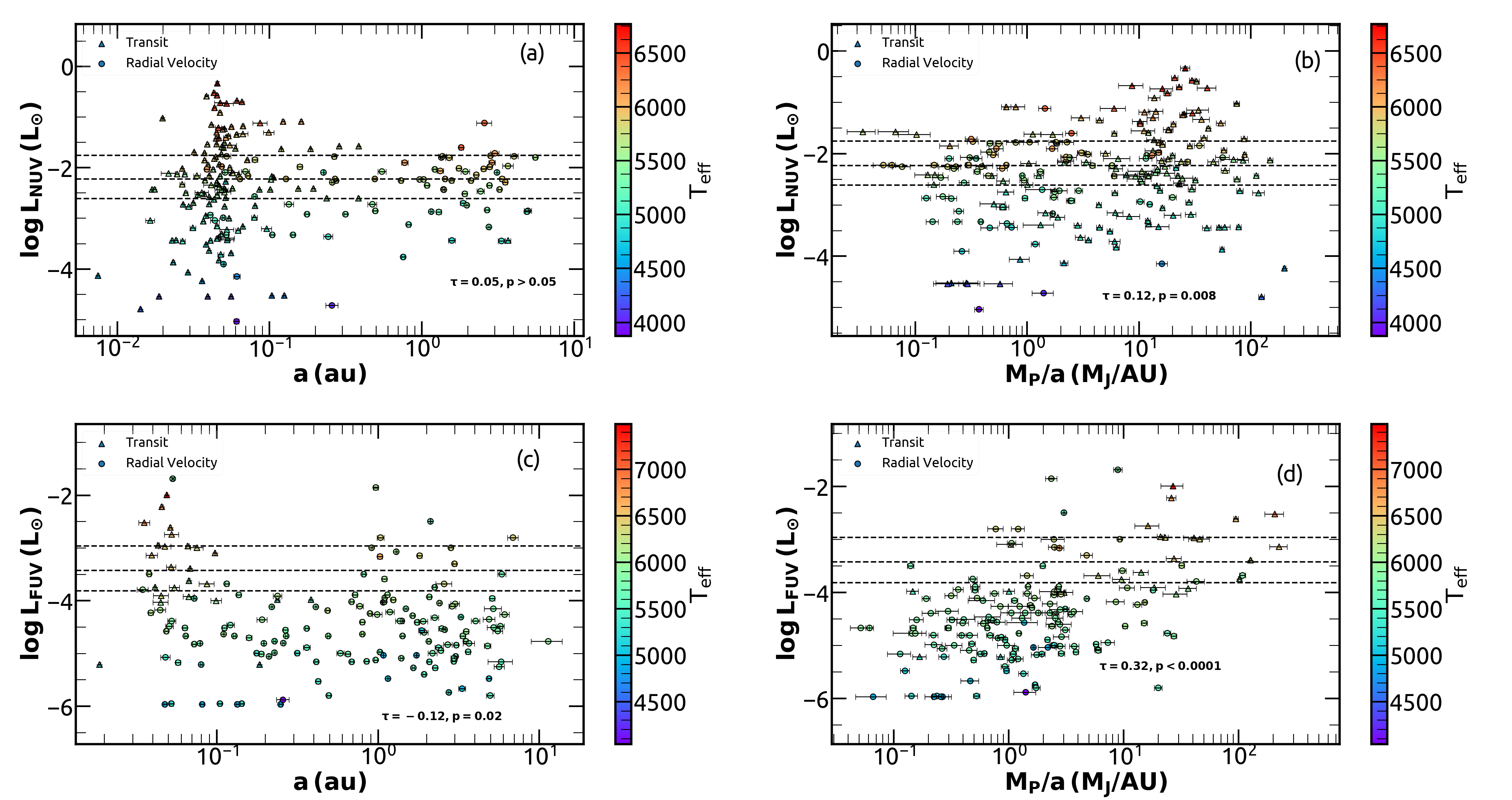}
\caption{Panels (a),(b) shows the variation of $\mathrm{L_{NUV}}$ of the main sequence host stars with their planetary properties $a$ and \mp/$a$. Panels (c),(d) shows the variation of $\mathrm{L_{FUV}}$ of the main sequence host stars with $a$ and \mp/$a$. Here the radial velocity detected planets are shown as solid circles and the transit detected planets are shown as diamonds. Each source is color-coded according to the \teff~ of the host star. The black dashed line indicate the lower quartile, median and upper quartile of y-axis values for the non-planet hosting $Gaia$ sample. The Kendall tau correlation coefficients and the corresponding p-values are also shown in each plot.} \label{Fig2}
\end{figure*}

\subsection{Variation of host star UV activity with planet properties}

Figure \ref{Fig2} shows the variation of the NUV and FUV luminosity of the main sequence host stars with the planetary properties, with the Radial velocity and Transit detected planets shown separately. We see no strong correlation between the NUV luminosity of these main sequence host stars and their planetary properties (Figure \ref{Fig2}(a), \ref{Fig2}(b)). We tested for correlation among these properties using the Kendall Tau correlation test, which is a non-parametric test that measures the strength of statistical association between two variables using the correlation rank coefficient and an associated probability that two randomly drawn samples of these variables may produce this correlation. The two variables are considered to have a strong positive correlation if the rank coefficient is close to 1 (or $-1$ for negative correlation) and the p-value is close to 0. The Kendall Tau rank coefficient between $\mathrm{L_{NUV}}$ and $a$ has a value of 0.05, with an associated probability of p $>$ 0.05. Similarly, the Kendall Tau coefficients between $\mathrm{L_{NUV}}$ and \mp/$a$ is 0.12 (p = 0.008). The FUV luminosity of the host stars also do not show any scientifically significant correlation with their planetary orbital distance, as seen in Figure \ref{Fig2}(c) (Kendall's tau = $-0.12$, p = 0.02). On the other hand, the Kendall's tau correlation test for Figure \ref{Fig2}(d) between $\mathrm{L_{FUV}}$ and \mp/$a$ does hint at some correlation, with $\tau$ = 0.32 and p $<$ 0.0001. However, this correlation is likely due to two reasons that do not point to SPI: a) the small number of Transit detected planets in the FUV sample, b) the contribution from the photosphere of the host star. In the Radial Velocity method of planet detection, the stellar activity is potentially a source of noise. Hence, around more active stars, the RV method tends to detect only those planets having higher \mp/$a$. In the Transit method of planet detection, however, the stellar activity introduces less detection bias. With the FUV sample predominantly consisting of RV detected planets as compared to its small number of Transit detected planets, it will on an average show an increase in \mp/$a$ with increasing FUV luminosity, consequently resulting in the correlation in Figure \ref{Fig2}(d). To explore this further, we extended the number of Transit detected planets in our FUV sample by including those planets for which planet mass was not listed in the Confirmed Planets table. For these planets, we calculated their planet mass from their respective planet radii using the \cite{CH2017} relation as described in the Composite Planet Data table. This resulted in the addition of 10 more transit detected planets in our FUV sample, giving a sample size of 177 planets around 133 FUV detected host stars. With this new sample, we repeated the Kendall's tau correlation test between $\mathrm{L_{FUV}}$ and \mp/$a$ and found that the slight increase in the sample size led to a drop in the correlation strength to $\tau$ = 0.24 from 0.32, with p $<$ 0.0001. This hints that the less number of Transit detected planets in the sample definitely has a significant role in driving the correlation in Figure \ref{Fig2}(d).

More importantly, the NUV as well as FUV luminosity of stars have an intrinsic dependence on the stellar surface temperature, which results in considerable contribution from the star's photosphere to its UV luminosity. To understand this better, we color-coded the plots in Figure 2 according to the surface temperature of the host stars. As is evident from the figure, $\mathrm{L_{NUV}}$ as well as $\mathrm{L_{FUV}}$ show strong correlation with stellar \teff~ (The Kendall Tau rank coefficient between $\mathrm{L_{NUV}}$ and \teff~ has a value of 0.74 with p$<$0.0001 and between $\mathrm{L_{FUV}}$ and \teff~ has a value of 0.63 with p$<$0.0001). Although for the NUV sample, the orbital distance $a$ and the SPI indicator \mp/$a$ do not show any practically significant correlation with the stellar \teff~, for the FUV sample \mp/$a$ shows a positive correlation with \teff~ ($\tau=0.32;~p<0.0001$), indicating that planets with higher \mp/$a$ are generally associated with hotter, hence more FUV and NUV luminous stars (similar to the conclusions drawn in \cite{France2018}). This effectively results in a positive trend between $\mathrm{L_{FUV}}$ and \mp/$a$. Such trends that result from underlying correlations with stellar parameters could be mistaken as an indication of SPI-related activity. Thus, while searching for SPI signatures it is important to account for this photospheric contribution to the UV activity. Since most of our host stars are FGK type stars, we use the  MESA Isochrones and Stellar Tracks (MIST\footnote{http://waps.cfa.harvard.edu/MIST/}; \cite{Choi2016}) isochrone for dwarfs corresponding to 1 Gyr to model the NUV and FUV contribution of the photosphere to the stellar activity (overplotted in Figure \ref{Figure1}). From the MIST isochrones, we then calculate the model log($\mathrm{L_{NUV}/L_{bol}}$) and log($\mathrm{L_{FUV}/L_{bol}}$) values for the host stars in our NUV and FUV samples respectively. These values are then subtracted from the observed values of log($\mathrm{L_{NUV}/L_{bol}}$) and log($\mathrm{L_{FUV}/L_{bol}}$) of the host stars, to obtain the excess fractional NUV and FUV luminosity values:

\begin{equation}
\begin{split}
    \mathrm{\Delta log(L_{NUV}/L_{bol}) = log(L_{NUV}/L_{bol})_{obs} -}\\ \mathrm{log(L_{NUV}/L_{bol})_{MIST}}\\
    \mathrm{\approx log(L_{{NUV}_{obs}}/L_{{NUV}_{MIST}})} \\
    \text{\small{(Assuming similar predicted and observed $\mathrm{L_{bol}}$)}}
\end{split}
\end{equation}

\begin{equation}
\begin{split}
    \mathrm{\Delta log(L_{FUV}/L_{bol}) = log(L_{FUV}/L_{bol})_{obs} -}\\ \mathrm{log(L_{FUV}/L_{bol})_{MIST}}\\
    \mathrm{\approx log(L_{{FUV}_{obs}}/L_{{FUV}_{MIST}})} \\
    \text{\small{(Assuming similar predicted and observed $\mathrm{L_{bol}}$)}}
\end{split}
\end{equation}

\begin{figure*}[ht!]
\centering
\includegraphics[width=1.0\linewidth]{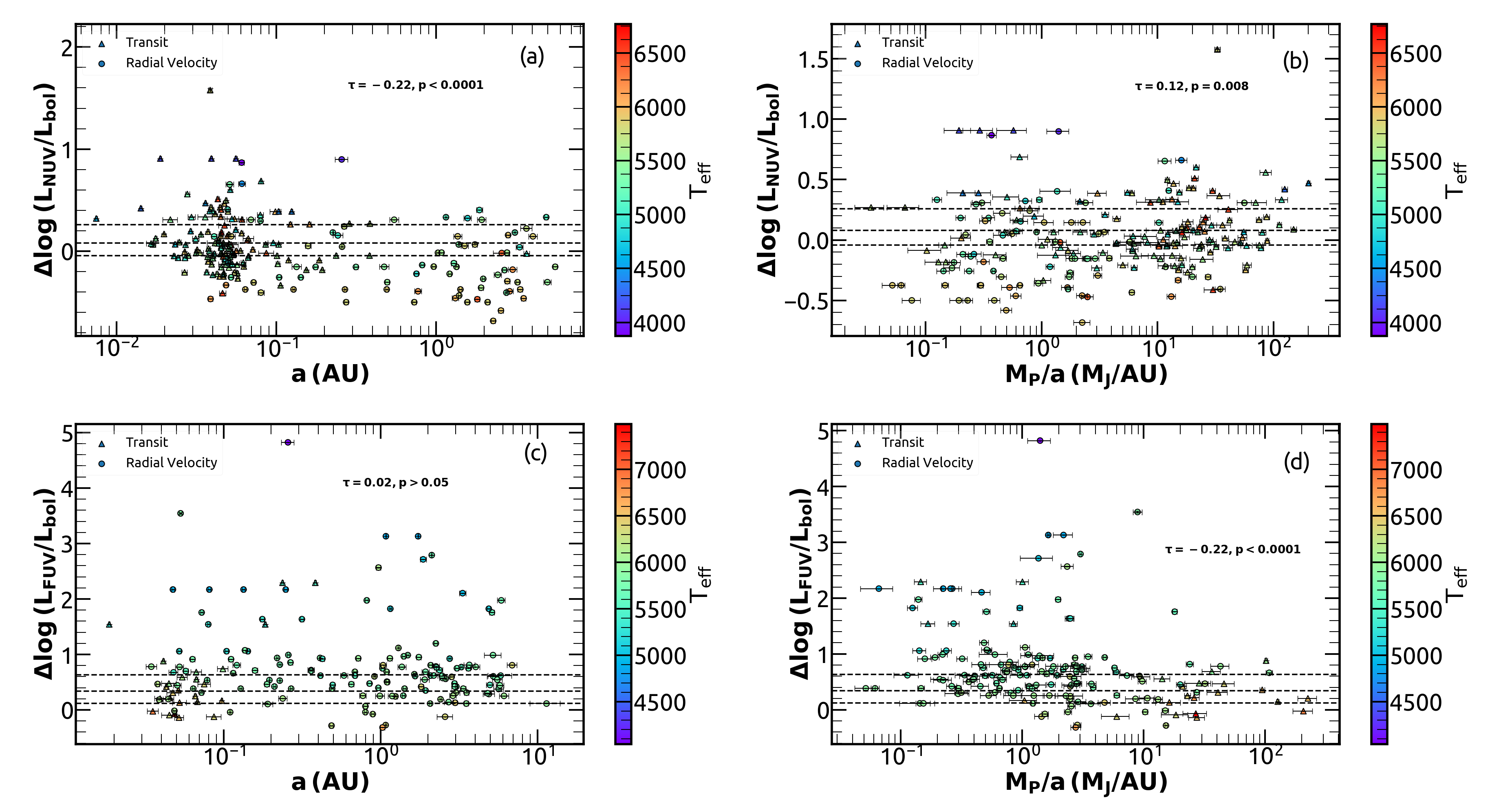}
\caption{Panels (a),(b) show the variation of excess fractional NUV luminosity of the main sequence host stars with $a$ and \mp/$a$. Panels (c),(d) show the variation of excess fractional FUV luminosity of the main sequence host stars with $a$ and \mp/$a$. The colours and symbols hold the same meaning as in previous figure.} \label{Fig3}
\end{figure*}

The excess fractional luminosity obtained here thus expresses the observed fractional luminosity of the host star as a fraction of the luminosity due its photospheric emission as predicted from the model, in log scale, thus effectively removing the contribution of photosphere to the fractional luminosity. Further, if the $\mathrm{L_{bol}}$ values from the observations and the MIST model are similar, this quantity simply gives the observed UV luminosity as a fraction of the model predicted UV luminosity in log scale. We then plotted the excess fractional NUV and FUV luminosity of the main sequence host stars against the planetary properties as shown in Figure \ref{Fig3}. Here, unlike in Figure \ref{Fig2}, the data points color-coded according to the stellar \teff~ no longer show a trend of NUV and FUV luminosity with \teff.     

The excess fractional NUV luminosity in Figure \ref{Fig3}(a) shows a weak negative correlation with the orbital distance, with the Kendall Tau rank coefficient between $\mathrm{\Delta log(L_{NUV}/L_{bol}})$ and $a$ being $-0.22$ and p $<$ 0.0001. From Figure \ref{Fig3}(b), we see that $\mathrm{\Delta log(L_{NUV}/L_{bol})}$ shows no practically significant correlation with  \mp/$a$ ($\tau$ = 0.12, p = 0.008). Figure \ref{Fig3}(c) and \ref{Fig3}(d) analyze the variation of excess fractional FUV luminosity with the same planetary properties. We do not find any significant correlation between $\mathrm{\Delta log(L_{FUV}/L_{bol})}$ and $a$ , with the Kendall Tau rank correlation coefficient being 0.02 (p $>$ 0.05). For Figure \ref{Fig3}(d), we see a weak negative correlation between $\mathrm{\Delta log(L_{FUV}/L_{bol})}$ and \mp/$a$ ($\tau = -0.22$, p $<$ 0.0001), similar to Figure \ref{Fig3}(a). We see here that the positive correlation found in Figure \ref{Fig2}(d) between $\mathrm{L_{FUV}}$ and \mp/$a$ is no longer present in Figure \ref{Fig3}(d), once we remove the photospheric contribution. It is now evident that the positive correlation seen in Figure \ref{Fig2}(d) is driven by the photospheric emission from the host star and hence do not constitute evidence of SPI. However, the negative correlation seen in Figures \ref{Fig3}(a) and \ref{Fig3}(d) needs to be further explored. Here, the apparent trend seems to be mainly driven by the distinct nature of radial velocity detected planets in both the cases. To confirm this, we extended the number of Transit detected planets in our NUV sample in the same way how we extended our FUV sample, by calculating the planet's mass from its radius for those planets whose mass was not listed in the Confirmed Planets table. This increased the number of transit detected planets in NUV sample resulting in a sample size of 515 planets around 413 NUV detected host stars. Using this larger sample, we calculated the Kendall Tau correlation coefficients for Figures \ref{Fig3}(a) and \ref{Fig3}(d). We found that the correlation strength is reduced in both the cases, with $\tau~ =~-0.07$ (p = 0.02) between $\mathrm{\Delta log(L_{NUV}/L_{bol})}$ and $a$ and $\tau=-0.18$ (p = 0.0003) between $\mathrm{\Delta log(L_{FUV}/L_{bol})}$ and \mp/$a$. Note that while for the former case the correlation is greatly weakened, for the latter case due to the lesser number of transit detected planets around FUV host stars even on extending the sample, the reduction in correlation strength is not much, however a definite weakening of the apparent correlation in this case is reassuring. Thus, we can safely say that the negative correlation found in Figures \ref{Fig3}(a) and \ref{Fig3}(d) are driven by the distinct nature of Transit and Radial Velocity detected planets and are likely due to detection-bias rather than any SPI-related activity. However, unlike \cite{Shkolnik2013}, we do not find any statistically significant evidence for higher excess fractional FUV luminosity in the host stars with close-in planets as compared to the ones with far out planets.

To confirm the behaviour seen from Figure \ref{Fig3}, we also obtained the excess fractional luminosity of the host stars via an alternate method. We fit a second order polynomial function to the Gaia sample of non-planet hosting main sequence stars within 100 pc to predict the fractional NUV, FUV luminosity for a given stellar surface temperature. In order to avoid outliers from the fit, we restricted the sample size using 3$\sigma$ clipping technique and obtained the following polynomial fit between log fractional luminosity and surface temperature for NUV and FUV (as shown in Figure \ref{Figure1}):

\begin{equation}
    \begin{split}
               \mathrm{log~(L_{FUV}/L_{bol})} = 7.6\times 10^{-7} ~\mathrm{T_{eff}^2}~-~8.3\times 10^{-3}~\mathrm{T_{eff}}\\+~18.24
    \end{split}
\end{equation}

\begin{equation}
    \begin{split}
               \mathrm{log~(L_{NUV}/L_{bol})} = -4.9\times 10^{-7} ~\mathrm{T_{eff}^2}~+~6.3\times10^{-3}~\mathrm{T_{eff}}\\-~22.25
    \end{split}
\end{equation}

\begin{figure}[ht!]
\centering
\includegraphics[width=1.0\linewidth]{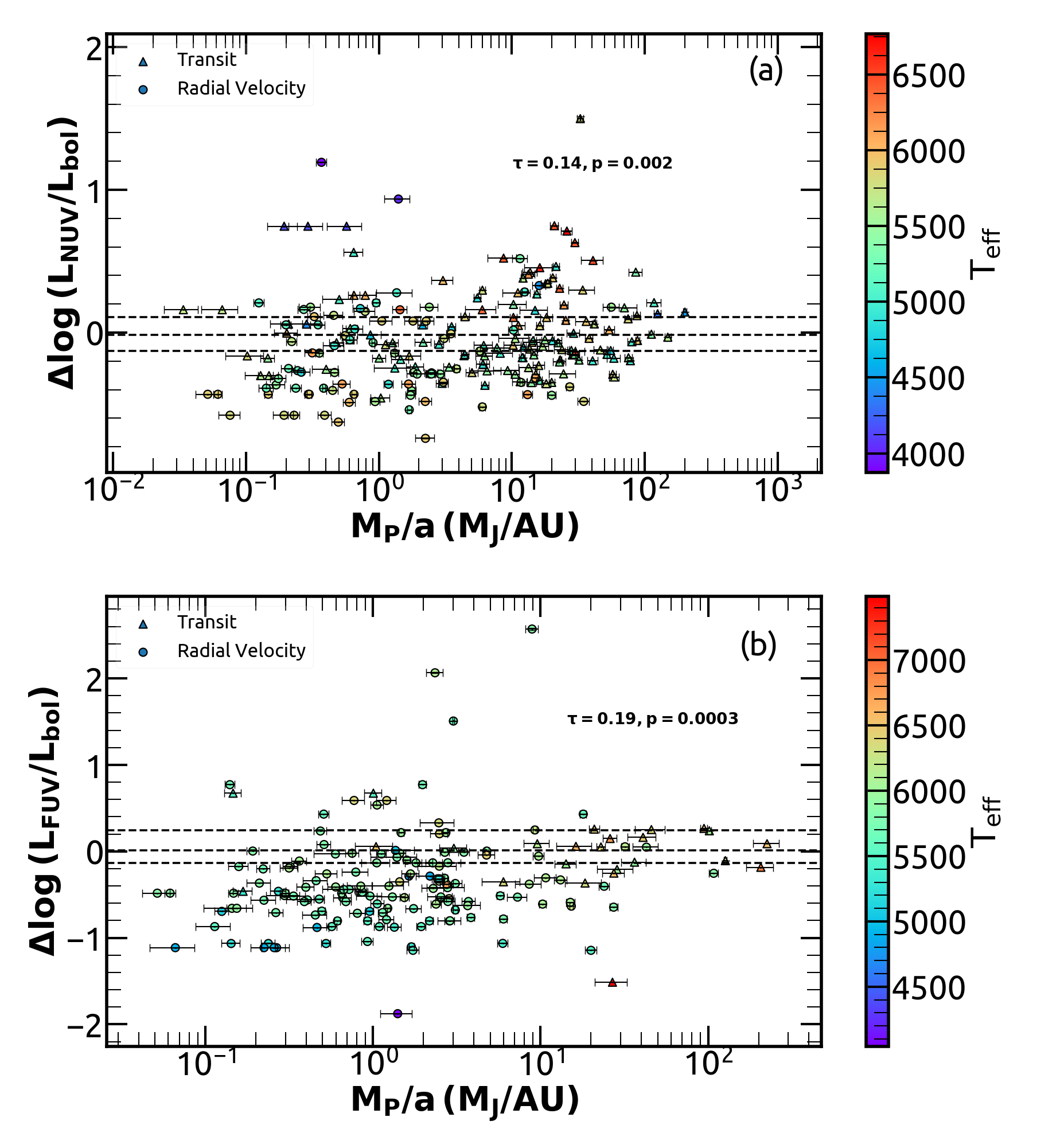}
\caption{The variation of excess (a) NUV and (b) FUV luminosity of the main sequence host stars obtained using the polynomial fit to the {\it Gaia} sample of non-planet hosting main sequence stars within 100 pc shows no practically significant correlation with \mp/$a$. The colours and symbols hold the same meaning as in previous figures.} \label{Fig4}
\end{figure}

The log fractional luminosity thus predicted was then subtracted from the log of observed fractional luminosity, to obtain the excess fractional NUV, FUV luminosities of the host stars. The variation of these with \mp/$a$ (as shown in Figure \ref{Fig4}) also show similar behaviour as with the analysis using MIST isochrone, further confirming the absence of any correlation between the NUV, FUV activity of the host stars and planet properties.

Since most of the SPI reported in literature were in F, G, K type stars and also since SPI is mainly induced due to massive hot Jupiters, in order to carry out a more careful inspection of our results, we did a similar analysis using (a)  a sample of only those main sequence host stars with F, G, K spectral types, (b) a sample of only massive planets (\mp $>$ 100 $M_{\oplus}$), (c) a sample of only close-in planets ($a<=0.1$ au) and (d) a sample of only massive close-in planets (\mp $> 100 M_{\oplus}$ and $a<=0.1$ au). However, the results obtained were similar to that of the larger sample, and there was no significant correlation found between the stellar luminosities and the planetary properties. Any small correlation that emerged between the UV luminosities and $a$ or \mp/$a$ in any of these cases disappeared once the photospheric contribution was removed, similar to the case with the parent sample.

\subsection{Comparing the host star distribution with that of known chromospherically active stars}

Since we do not find any statistical correlation from our analysis that is indicative of SPI, we next check whether broad band UV flux effectively traces chromospheric activity by comparing the UV luminosity of host stars in our sample to that of chromospherically active stars. For this, we used the sample of stars detected in RAVE DR5 \citep{Kunder2017} from \cite{Zerjal2017}. The RAVE (RAdial Velocity Experiment) survey uses the Ca\textsc{ ii} infrared triplets (Ca\textsc{ ii} IRT) at 8498, 8542 and 8662 \AA~to study the chromospheric activity of stars. The activity identification was done by identifying excess emission flux in the Ca\textsc{ ii} IRT while the rest of the spectrum remains indistinguishable from an inactive state. Photospheric component of the flux was eliminated from the active candidates by subtracting the best-matching inactive template spectrum. The activity proxy used is the combined Equivalent Width (EW$\mathrm{_{IRT}}$) of each calcium line in the spectrum of the star:

\begin{equation}
    \mathrm{EW_{IRT}} = \mathrm{EW_{8498}} + \mathrm{EW_{8542}} + \mathrm{EW_{8662}}
\end{equation}

\cite{Zerjal2017} found the distribution of EW$\mathrm{_{IRT}}$ of inactive stars to be centred around $-0.05$ \AA~with a $\sigma$=0.16 \AA~and hence used 0.16 \AA~as the activity detection limit. Subsequently, out of the 38678 candidate spectra they catalogued, they reported about 13,000 stars with activity level above 2$\sigma$ and about 22,000 stars with activity above 1$\sigma$ respectively.

We cross-matched all the stars from \cite{Zerjal2017} catalogue having spectral SNR above 50 with the GALEX GR6/GR7 and $Gaia$ DR2 catalogues to obtain a sample of 7124 NUV and 652 FUV stars with good GALEX photometry and the Gaia stellar parameters measured above 3$\sigma$ accuracy.

\begin{figure*}[ht]
\centering
\includegraphics[width=1.0\linewidth]{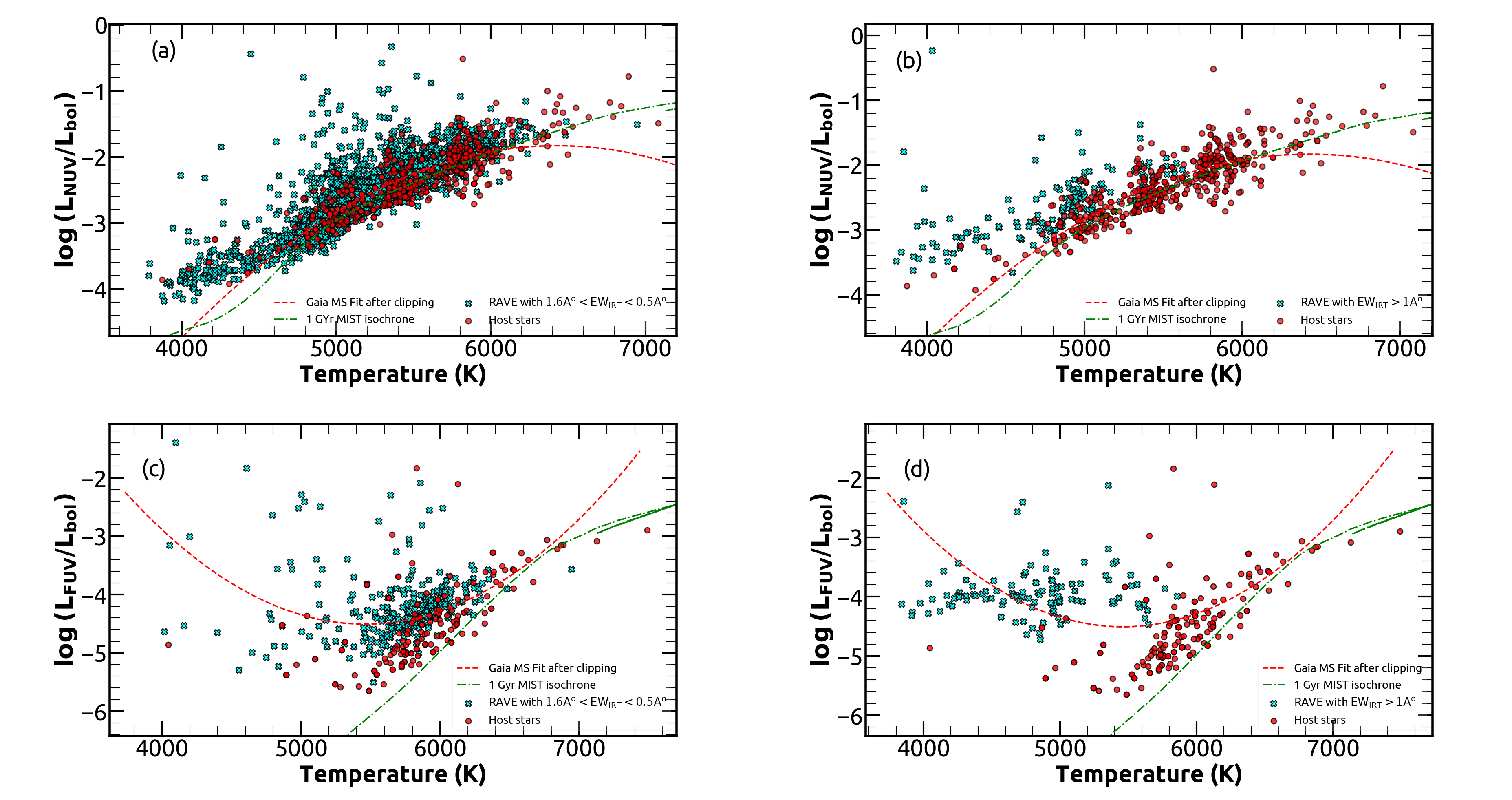}
\caption{The distribution of the GALEX and {\it Gaia} detected RAVE sample from \cite{Zerjal2017} in the fractional NUV/FUV luminosity vs $\mathrm{T_{eff}}$ plane. The cyan solid crosses indicate the RAVE stars and the red solid circles indicate the main sequence host stars. Panels (a) and (c) consists of moderately active RAVE stars with $\mathrm{EW_{IRT}}$ between 0.16 and 0.5 \AA~ while panels (b) and (d) consists of highly active RAVE stars with $\mathrm{EW_{IRT}}>$~1~\AA~.} \label{Fig5}
\end{figure*}

Figure \ref{Fig5} shows the distribution of the GALEX and Gaia detected RAVE sample in the fractional NUV/FUV luminosity vs \teff~ plane, separately analyzed on the basis of activity levels. From Figures \ref{Fig5}(a) and \ref{Fig5}(b), we see that while the low and moderately active RAVE stars (0.16 \AA~$\mathrm{>EW_{IRT}>}$~0.5 \AA) have a similar distribution to that of main sequence host stars, the highly active RAVE stars ($\mathrm{EW_{IRT}>}$~1~\AA) occupy a slightly higher distribution to that of the host stars in the $\mathrm{L_{NUV}/L_{bol}}$ vs $\mathrm{T_{eff}}$ plane. This distinction is much more evident in FUV (Figures \ref{Fig5}(c) and \ref{Fig5}(d)). This result indicates that any enhancement in the chromospheric activity of host stars due to SPI may not be high enough to cause a significant increase in their UV luminosity comparable to highly active stars. 

\begin{figure}[ht!]
\centering
\includegraphics[width=1.12\linewidth]{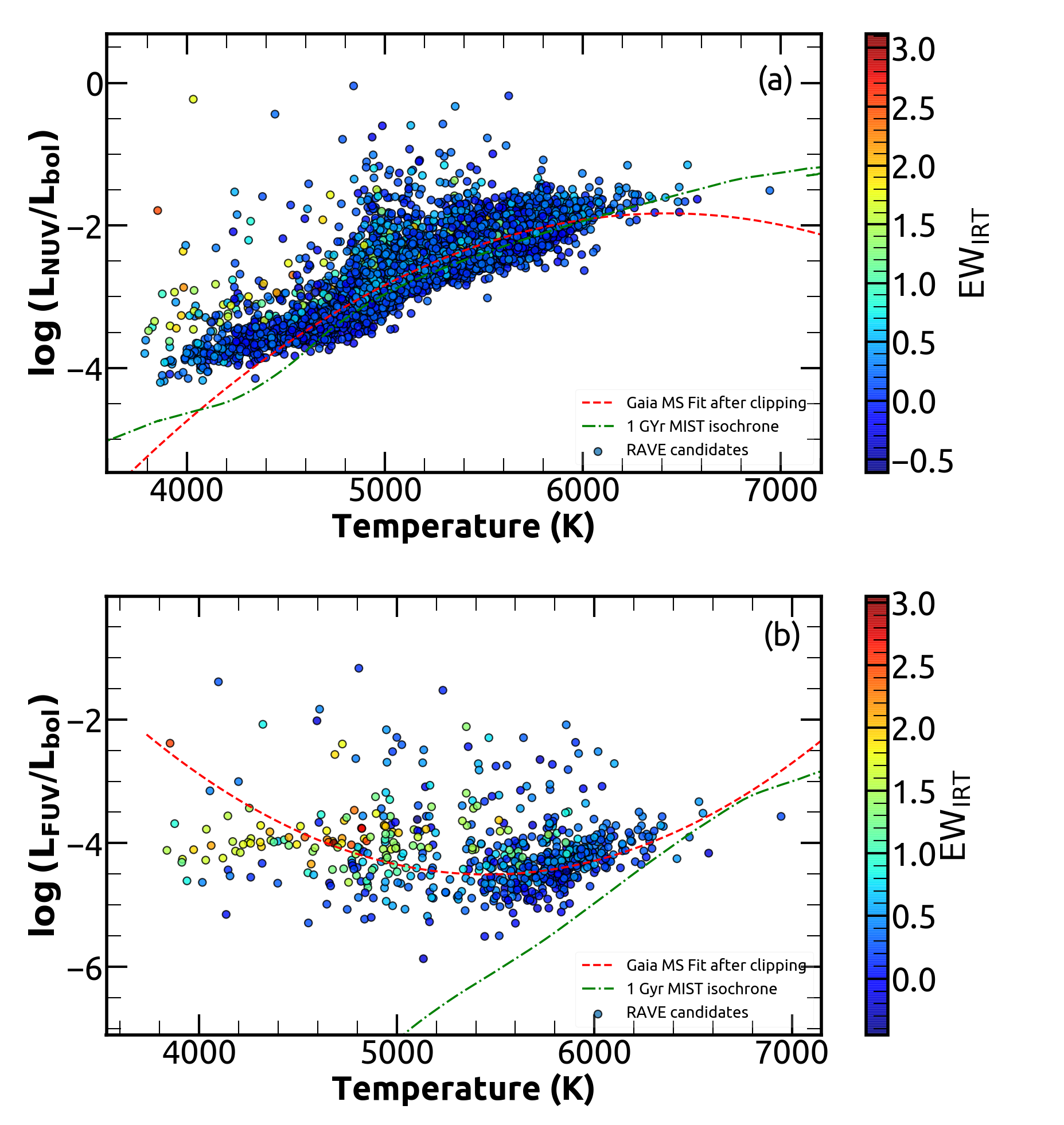}
\caption{The variation of (a) fractional NUV luminosity and (b) fractional FUV luminosity of the GALEX and {\it Gaia} detected RAVE sample from \cite{Zerjal2017}. The data points are color-coded according to their activity indicator ($\mathrm{EW_{IRT}}$). The distribution of the {\it Gaia} detected non-planet hosting dwarfs is indicated by the polynimial fit, similar to Figure \ref{Figure1}.} \label{Fig6}
\end{figure}

To further investigate this, we carefully analyzed the $\mathrm{EW_{IRT}}$ of the active RAVE stars with respect to their UV luminosity. Figure \ref{Fig6} shows the fractional NUV/FUV luminosity vs $\mathrm{T_{eff}}$ plots for these samples, color-scaled according to their activity ($\mathrm{EW_{IRT}}$). From their analysis using [J-K] vs. [NUV-V] diagram, \cite{Zerjal2017} found that even though the moderately active stars have colors similar to inactive stars, the most active stars (EW$\mathrm{_{IRT}}$ $>$~1 \AA) are significantly bluer in [NUV-V] compared to the inactive stars. Along similar lines, Figure \ref{Fig6} also does not show any distinction between the distributions of stars with low and medium chromospheric activity in the luminosity-temperature plot, both in NUV as well as FUV. Only the highest activity stars (EW$\mathrm{_{IRT}}$ $>$~1~\AA) lie distinctly above the rest of the sample as well as above the {\it Gaia} main sequence distribution. 

\begin{figure}
\centering
\includegraphics[width=1.12\linewidth]{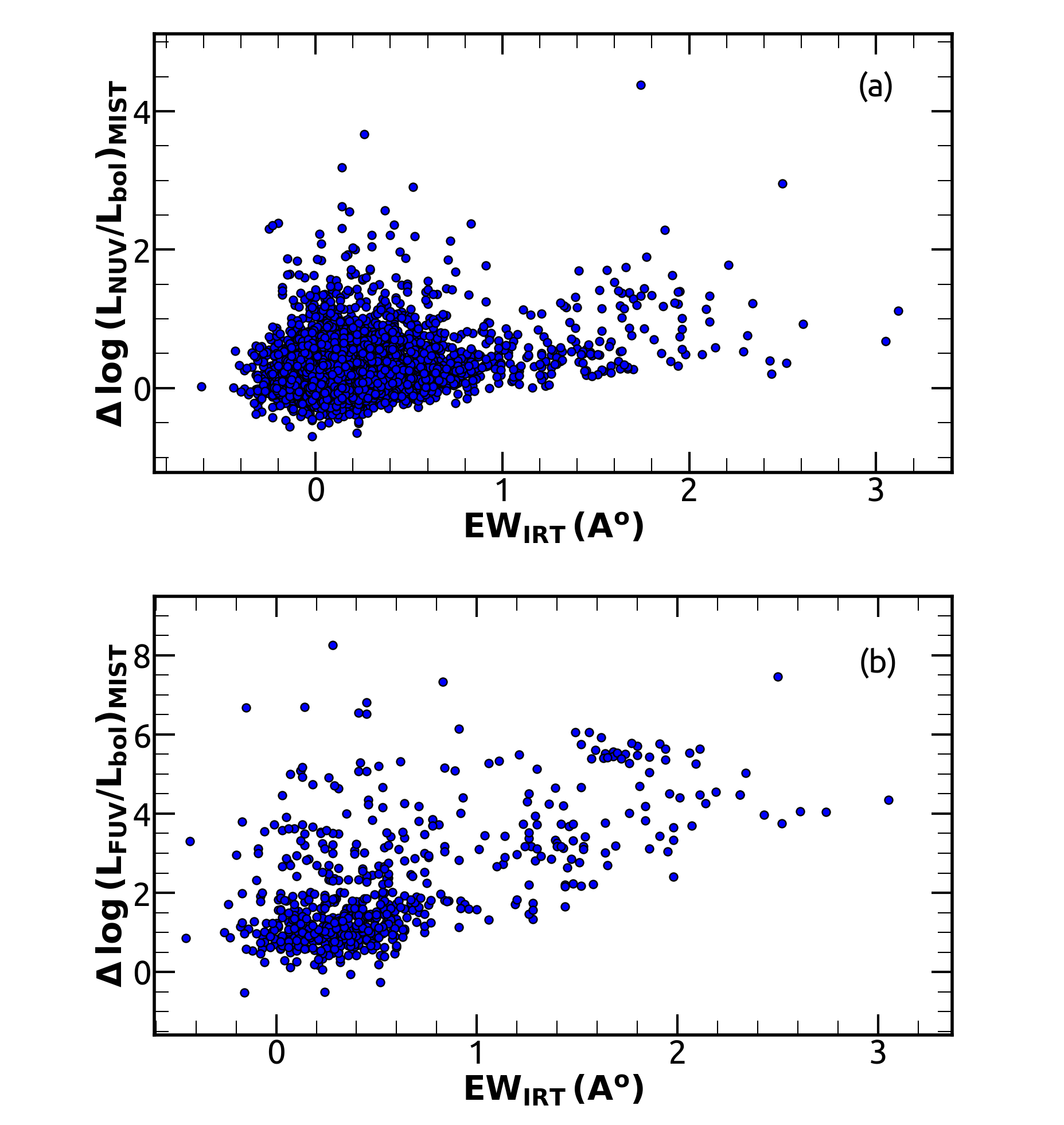}
\caption{The variation of (a) excess fractional NUV luminosity and (b) excess fractional FUV luminosity with $\mathrm{EW_{IRT}}$ for the GALEX and {\it Gaia} detected RAVE sample from \cite{Zerjal2017}.} \label{Fig7}
\end{figure}

This absence of a trend in UV for the medium to low-level chromospheric activity of stars is clearer when we look at the variation of excess fractional luminosity for these RAVE stars (ie. after subtracting the photospheric contribution in UV luminosity using 1 Gyr MIST isochrones) with the activity proxy (Figure \ref{Fig7}). Here we do see a weak correlation between the excess fractional UV luminosities of the overall sample with their $\mathrm{EW_{IRT}}$; Between $\mathrm{\Delta log(L_{NUV}/L_{bol})}$ and $\mathrm{EW_{IRT}}$, the Kendall's Tau = 0.11 (p $<$ 0.0001) and between $\mathrm{\Delta log(L_{FUV}/L_{bol})}$ and $\mathrm{EW_{IRT}}$, the Kendall's Tau = 0.34 (p $<$ 0.0001). In these cases, although the correlation strength is very small, the probability values are statistically significant. However, this correlation is mainly driven by the highly active stars. On separately analyzing just those stars with $\mathrm{EW_{IRT}}<$ 1~\AA~, the correlation strength significantly decreases; between $\mathrm{\Delta log(L_{NUV}/L_{bol})}$ and $\mathrm{EW_{IRT}}$, the Kendall's Tau = 0.08 (p $<$ 0.0001) and between $\mathrm{\Delta log(L_{FUV}/L_{bol})}$ and $\mathrm{EW_{IRT}}$, the Kendall's Tau = 0.15 (p $<$ 0.0001). Hence, using the RAVE sample of active stars, we see that chromospheric activity when measured using the  Ca\textsc{ ii} IRT shows a correlation with their GALEX UV broadband flux only for highly active stars, but for low-medium activity stars, UV broadband flux do not seem to effectively trace the chromospheric activity. This lack of any significant correlation among the low-medium active stars could also be due to the fact that GALEX measurements of NUV, FUV flux and the RAVE measurements of Ca\textsc{ ii} IRT flux are not taken during the same epoch. Differences in stellar activity could occur between the two epochs caused due to factors like stellar rotation, activity cycle etc., explaining the non-correlation. Nevertheless, this analysis does point to the possibility that activity in the upper stellar atmospheres actually starts showing up in GALEX UV bands only for the highly active stars with certainty.

\section{Discussion}

As discussed in Section 1, the activity in the upper stellar atmospheres can have considerable effects in the UV region of the electromagnetic spectrum. Moreover, from our analysis of Figure \ref{Fig6} and Figure \ref{Fig7} in Section 3.3, we see that broad band UV flux effectively traces the chromospheric activity of highly active stars, which provides a reasonable incentive to search for enhanced UV activity among the planet hosting stars due to SPI. In our work, we looked for signatures of possible Star-Planet Interaction in the UV activity of the host stars  statistically by studying their latest GALEX NUV and FUV luminosities against their planetary properties, mainly orbital distance and the ratio of planetary mass to orbital distance. Our analysis indicate that there is no significant correlation between the stellar UV luminosity and the associated planetary properties that could be indicative of an SPI signature in UV.

\vspace{2mm}
\subsection*{W\MakeLowercase{hy do we not find a statistical} SPI \MakeLowercase{signature in} UV?}\vspace{3mm}

These could be some possible explanations for the absence of evidence for SPI in our analysis.

Firstly, the GALEX images and UV magnitudes of the host stars are simply snapshots, or in other words, single epoch measurements. Stellar activity itself is a time-variable phenomenon. To add to it, as was mentioned in Section 1, there is just roughly 75\% probability that the SPI signatures may show up. Perhaps a time resolved analysis of the host star's UV flux would be a better indicator of the variation in their chromospheric activity due to associated planetary properties. 

Another possible reason for the lack of statistical evidence of an SPI signature from our analysis could simply be that the current sample of confirmed planetary systems do not yet have a large enough number of massive close-in planets capable of inducing such interactions with their host stars. 

A more significant reason could be that GALEX broadband UV flux is a good indicator of stellar activity in the upper atmospheres only for the highly active stars. Our analysis using the RAVE sample of active stars in Section 3.3 hints that the low to medium-level chromospheric activity measured in terms of Ca\textsc{ ii} IRT do not reflect well in the GALEX broad-band UV flux, showing that GALEX NUV and FUV flux may only indicate a star's chromospheric activity if it has a high activity-level. As is clearly seen from the the fractional UV luminosity vs \teff~ plot in Figure \ref{Fig5}, the separate distributions of the highly activity RAVE stars and the planet hosting stars indicate that even if SPI induces enhancement in UV activity, it may not be as high as the highly active stars detected in RAVE. This suggests that SPI induced enhancement in chromospheric activity may not be significant enough to be traced using UV broad band flux. This would make it really difficult to detect them in GALEX bands statistically, as these bands only reflect high-level activity. A possible caveat in the above analysis is that while majority of the highly active RAVE stars in FUV occupy a \teff~ range of 4000-5000 K, there are very few FUV detected host stars in this range. 

One concern regarding our analysis is the potential selection effects introduced in our sample by GALEX, which tends to detect more active stars. This would mean that selecting only GALEX-detected stars for our analysis would result in our sample generally having more active stars. However, if such a selection effect would have indeed been present, this would further add strength to our above analysis and subsequent conclusion, since  even among such active planet-hosting stars detected by GALEX we did not find evidence for an SPI signature. Further, GALEX measures the flux from the star, hence the distance to the star is an important factor. Apart from the active stars, stars that may be less active but are closer to us may also be detected by GALEX. However, GALEX will not be able to detect very distant active stars. This will only affect our results if the SPI-induced activity enhancement occurs only for those stars that are beyond a certain distance. This scenario would be highly unlikely.  

Although there have been claims in the literature about the presence of a statistical evidence for SPI signature in Ca\textsc{ ii} K line \citep{KB2012}, UV \citep{Shkolnik2013} and X-ray \citep{Kashyap2008}, recent works point to ambiguity regarding the statistical significance of SPI. \cite{Popp2010} studied the X-ray and fractional X-ray luminosities of a sample of planet hosting stars and found no significant correlation with the interaction proxy \mp/$a$. \cite{Canto2011} analyzed the activity of 74 planet hosting stars and found no significant correlation between the activity indicator log(R'$\mathrm{_{HK}}$) and the planetary properties $a$ and \mp/$a$. Similar to \cite{Popp2010}, they attribute any possible trend to selection effects. \cite{Miller2012} studied the WASP-18 planetary system and demonstrated that if Ca\textsc{ ii} H \& K variability is studied over just a short part of the rotation period, the observations can mistake a stellar hotspot for planet-induced activity. Further, \cite{Miller2015} analyzed a sample of planet hosting stars to find no statistical correlation between planetary properties and Ca\textsc{ ii} H \& K emission, in addition to finding no correlation between the SPI interaction proxy and X-ray luminosity of the sample. \cite{France2018} explicitly studied the FUV (1150 – 1450 \AA) emission lines (C\textsc{ ii}, Si\textsc{ iii}, Si\textsc{ iv} and N\textsc{ v}) of 71 planet hosting stars using HST-COS observations. While they do find statistically significant correlations between the UV activity levels of the host stars and \mp/$a$ of their planets, deeper analysis of these results using Principal Component Analysis to include the underlying correlations with stellar parameters revealed that SPI does not play a strong role in influencing the UV activity levels of their sample. This conclusion has been reinforced by \cite{Route2019} who studied HD189733 system in X-ray, UV, Ca\textsc{ ii} H \& K, H$\alpha$ and radio wavelengths and demonstrated that the stellar activity enhancements in this system are a likely result of inadequately sampled intrinsic stellar activity and not due to SPI as claimed by previous works.

\section{Conclusions}

We have used the largest sample of GALEX detected planet hosting stars with {\it Gaia} DR2 stellar parameters to look for correlation between their NUV/FUV activity and planetary properties to examine if star-planet interactions are detectable statistically in UV. From our analysis, we conclude the following:
\begin{itemize}
    \item We find no clear correlation between the NUV or FUV luminosity of main sequence planet hosting stars and their planetary orbital distance, $a$, or the ratio of planetary mass to orbital distance, \mp/$a$, that could be due to SPI-related activity.
    \item After removing the photospheric contribution to the UV flux using MIST isochrone as well as the fit to {\it Gaia} MS sample, we still find no correlation between the excess fractional NUV or FUV luminosity and the planetary properties $a$ and \mp/$a$, that could be due to SPI-related activity. Results from our analysis, while similar to those from the analysis in X-ray by \cite{Popp2010} and \cite{Miller2015}, are significantly different from the conclusions of \cite{Shkolnik2013}.
    \item Comparative analysis of the distributions of RAVE detected stars of various activity levels with the distributions of planet-hosting stars indicate that SPI induced enhancement in stellar activity, if any, may not be high enough to cause a significant increase in their UV luminosities comparable to highly active stars. 
    \item Analysis using RAVE detected chromospherically active main sequence stars indicate that the excess chromospheric activity measured via Ca\textsc{ ii} IRT lines only starts showing up in the GALEX broad band UV flux for the highly active stars ($\mathrm{EW_{IRT}}>$~1~\AA). This points to the possibility that if SPI-induced enhancement in chromospheric activity is modest, it may be difficult to detect them statistically using GALEX measurements.
\end{itemize}

\section{Acknowledgments}
We thank the referee for their insightful comments and suggestions that have led to the improvement of the final manuscript. This research is based on observations made with the GALEX mission, obtained from the MAST data archive at the Space Telescope Science Institute, which is operated by the Association of Universities for Research in Astronomy, Inc., under NASA contract NAS 5–26555. This research has made use of the NASA Exoplanet Archive, which is operated by the California Institute of Technology, under contract with the National Aeronautics and Space Administration under the Exoplanet Exploration Program.  This work presents results from the European Space Agency (ESA) space mission Gaia. Gaia data are being processed by the Gaia Data Processing and Analysis Consortium (DPAC). Funding for the DPAC is provided by national institutions, in particular the institutions participating in the Gaia MultiLateral Agreement (MLA). The lead author acknowledges the support from CHRIST (Deemed to be University) and the Visiting Student's Research Programme at Tata Institute of Fundamental Research, Mumbai, in undertaking this research work.

\end{document}